\documentclass[manuscript]{aastex}
\usepackage{tikz}
\usepackage{amsmath}

\usepackage{graphicx}

\usepackage{pdflscape}

\usepackage{url} 

\slugcomment{\textbf{Revised October 3} }

\shorttitle{Bennu}
\shortauthors{Jewitt}

\begin{document}

\title{Hubble Space Telescope Observations of \\
Active Asteroid (101955) Bennu}

\author{
David Jewitt$^{1}$, Max Mutchler$^{2}$, and Yoonyoung Kim$^1$
} 
\affil{$^1$Department of Earth, Planetary and Space Sciences, UCLA, 595 Charles Young Drive, Los Angeles, CA 90095\\}

\affil{$^2$Space Telescope Science Institute, Baltimore, MD 21218}

\email{djewitt@gmail.com}

\begin{abstract}
Measurements from NASA's OSIRIS-REx spacecraft revealed (101955) Bennu as a member of the active asteroid class.  These objects exhibit transient, comet-like activity, commonly producing  particle trails best observed from a vantage point close to the projected orbit plane.  Accordingly, we obtained deep, high resolution observations of Bennu at orbit plane crossing in 2023, using the Hubble Space Telescope.   We find no evidence for ejected material, and place model-dependent limits to the mass loss of $dM/dt < 0.12$ kg s$^{-1}$ from an analysis of the near-nucleus surface brightness profiles and $dM/dt < 6\times10^{-3}$ kg s$^{-1}$ from the absence of a measurable particle trail.  These limits are one to two orders of magnitude smaller than rates of mass loss measured remotely in other active asteroids and raise the possibility that low activity in many or most asteroids might fall beneath the prevailing  thresholds offered by remote observations.  Taking a still smaller  mass loss rate estimated from OSIRIS-REx data, we calculate a specific rate $\mathcal{L} \sim  10^{-7}$ [kg km$^{-2}$ s$^{-1}$] and extrapolate to the entire main-belt asteroid population.  We estimate a cumulative dust production rate $\sim850 r_{min}^{-1/2}$ [kg s$^{-1}$], where $r_{min}$ is the minimum asteroid radius in meters.  For $r_{min} >$ 1 m, this is $<$1\% to 10\% of the rate needed to sustain the Zodiacal cloud in steady state.
\end{abstract}


\section{INTRODUCTION}
\label{intro}
Asteroids and comets, formerly thought to be distinct products of  accretion on either side of the snow line, are now  recognized as end members of a continuum \citep{Jew24}.  In between the extremes, a few dozen bodies are known to have asteroidal orbits (Tisserand parameters with respect to Jupiter $T_J \gtrsim$ 3.08) while showing comet-like morphologies. These so-called ``active asteroids''  appear comet-like as a result of their intermittent ejection of dust. A surprising number of physical mechanisms behind the ejection have been identified.  Some active asteroids are indeed icy bodies (these are the ``main belt comets''; \cite{Hsi06}), in which subsurface ice sublimates when exposed by the sudden removal of a refractory debris mantle. Others are  produced by collisional shattering, impact ejection, thermal fatigue, and still others are products of rotational instability, driven either by torques from radiation or from anisotropic outgassing   \citep{Jew12}.

Observations from NASA's OSIRIS-REx mission show that near-Earth asteroid (101955) Bennu (formerly 1999 RQ36) is an active asteroid, episodically ejecting centimeter-sized particles  with speeds from a few cm s$^{-1}$ to $\sim$3 m s$^{-1}$ 
(\cite{Lau19a}, \cite{Lau19b}, \cite{Che20}).  
The process responsible for the ejection  remains unknown, with possible mechanisms being a subset of those already invoked to account for the active asteroids \citep{Jew12}. Specifically,  thermal fracture \citep{Mol20}, electrostatic launch \citep{Har22}, and impact of micrometeoroids \citep{Bot20} have all been considered as possible mechanisms on Bennu.  Ice sublimation seems unlikely given Bennu's proximity to the Sun and resulting mean blackbody temperature, T = 260 K, which is far too high for water ice or other common volatiles to persist near the surface.  

The orbit of  Bennu  has semimajor axis a = 1.126 AU, eccentricity e = 0.204 and inclination i = 6.0\degr, giving a Tisserand parameter with respect to Jupiter, $T_J$ = 5.525.  The perihelion lies interior to the Earth's orbit at q = 0.897 AU while aphelion, Q = 1.356 AU, is interior to the orbit of Mars and to the main asteroid belt from which Bennu is now dynamically decoupled.  Pre-spacecraft spectroscopic measurements indicated a B-type (i.e., optically blue) surface, suggesting minimal thermal alteration and the presence of hydrated minerals.  These properties, and easy access to Bennu from Earth-launched spacecraft, were the motivation for NASA's OSIRIS-REx mission.  Samples returned from this mission confirm  the  volatile rich nature and suggest a distant origin for Bennu, in the middle or outer solar system (\cite{Nut20}, \cite{Gla25}).

Bennu is  small, with an effective diameter of only $\sim$0.5 km \citep{Lau19a}. Sub-kilometer sized objects in the main asteroid belt have collisional lifetimes that are short compared to the age of the solar system, and their existence implies that they are fragments of a once larger body. Bennu is possibly  an escapee from the Polana-Eulalia collisional complex at $r_H \sim$ 2.4 au (\cite{Wal13}, \cite{Tak24}) and it is reportedly a second generation rubble pile, having been fragmented and re-formed at least once since its original collision \citep{Wal24}. Both the blocky surface and the internal porosity implied by Bennu's low bulk density ($\rho_B$ = 1190$\pm$15 kg m$^{-3}$; \cite{Goo21}) are  consistent with a rubble pile structure. In addition, the diamond-like or “top-shaped” body  of Bennu, produced by the equator-ward movement of material from the polar regions under centripetal acceleration, also reflects a fragmented and weak interior \citep{Rob21}. The present rotational period is 4.3 hours and, given its low density, Bennu remains close to rotational instability, where centripetal forces at the equator rival those of self-gravity.

The centimeter-sized particles detected at Bennu are individually 10 to 100 times larger (i.e., 10$^3$ to 10$^6$ times more massive) than the 0.1 mm to 1 mm particles inferred to dominate mass loss in most active asteroids. This difference is plausibly interpreted as an observational artifact because the cameras on OSIRIS-REx lack the sensitivity to detect particles much smaller than centimeter size. Conversely, small numbers of centimeter sized particles would not be detectable in ground-based observations, highlighting the mismatch between the sensitivity to asteroid activity of space-based and ground-based observations.   

\cite{Her20} concluded that material ejected from Bennu could not be detected in remote telescopic observations if the mass loss rate remains steady. However, observations of other active asteroids show that the mass loss is generally episodic, with order of magnitude variations occuring on timescales from days to years (\cite{Jew12}, \cite{Jew24}). They also show that debris trails, consisting of large, slow-moving particles, are most prominent when observed from close to the projected orbit plane.  In this paper we  analyze archival data and present new, highly sensitive, high resolution Hubble Space Telescope observations of Bennu taken from near the projected orbital plane.

\section{Observations}
Here, we consider three sets of  observations from the Hubble Space Telescope (HST).  First, we use archival observations from 2012 (Cycle 19, Program GO 13118, PI: M. Nolan).  Observations from this program were  taken to study the rotation of Bennu and have been used to obtain a 2.5$\sigma$ detection of rotational spin-up \citep{Nol19}. Observations under GO 13118 were obtained at two epochs, in 2012 September and 2012 December.  Second, we use new data from 2023 (Cycle 31, Program GO 17215, PI: Jewitt).  The observations were obtained on three dates (UT 2023 September 24, 25 and 26) but those on the 26th were found to have mis-pointed as a result of guide star acquisition failure, so only data from September 24 and 25 are useful.  All HST observations employed the WFC3 camera (image scale 0.04$\arcsec$ pixel$^{-1}$) and the F350LP filter, which provides maximum sensitivity to faint emission.  The latter is a long-pass filter with an extremely broad (equivalent width 4800\AA) passband, giving an effective wavelength near 6230\AA~when observing a source with a Sun-like spectrum.  

Table \ref{geometry} summarizes the geometry of observation on each date of observation. The observing geometries differ significantly between sets.  Notably, visits in 2023 September were timed to correspond to the crossing of the orbital plane of Bennu. Observations from this perspective provide maximum sensitivity to large particles which are confined by their low velocities to be close to the orbital plane. 

\subsection{Composite Images}

We reduced the data according to the following procedure.  Each image was visually examined to assess the impact of field stars and galaxies trailed relative to Bennu by the non-sidereal (largely parallactic) motion of the telescope.  Angular rates of Bennu peaked near 165\arcsec~hour$^{-1}$, corresponding to trail lengths $\sim$330 pixels in a  single 290 s image.
The most severely afflicted images were rejected or flagged so that their effect on the resulting image stacks could later be assessed.  We then shifted each image to a common center and rotated the data to align celestial North with the y-axis in our pixel array.  The images were then combined using a variety of statistical pixel rejection schemes.  In the end, we present unclipped median combinations of the images, since the median provides a simple and  robust measure of the distribution.

\begin{figure}
\epsscale{1.0}

\plotone{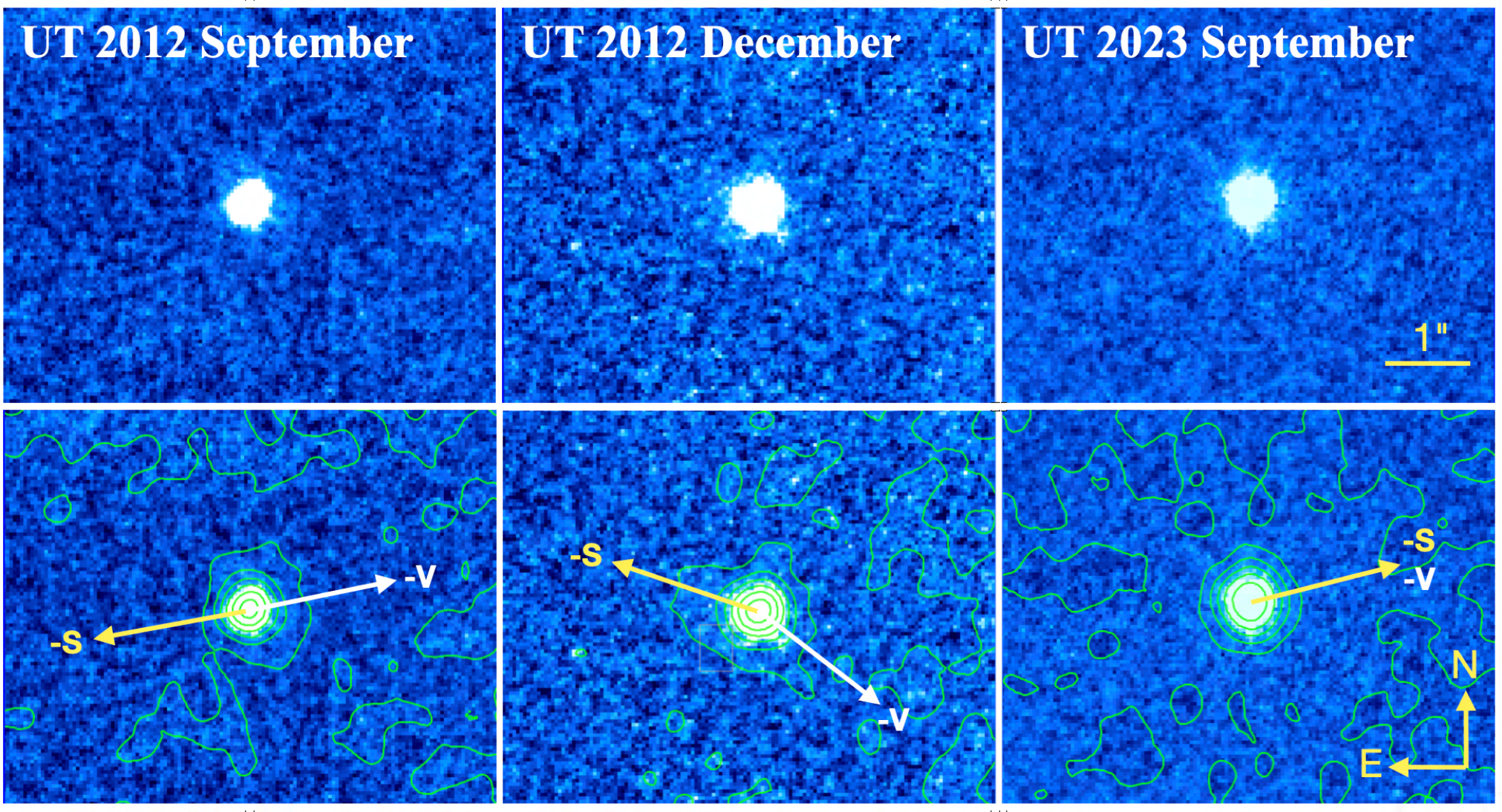}

\caption{(Top row:) Composite images on three dates (Table \ref{geometry}) shown with a common display range from -0.01 to 0.02 adu pixel$^{-1}$.  (Bottom row:) Same images from the top row contoured from 0.0004 adu pixel$^{-1}$ in steps of 0.002 adu pixel$^{-1}$.  Contours are smoothed over 8$\times$8 pixel boxes. Yellow arrows show the projected antisolar direction while white arrows show the projected negative heliocentric velocity vector, from Table \ref{geometry}.  The two directions are overlapping in the 2023 data.  A 1\arcsec~scale is common to all panels.   \label{composite}}
\end{figure}

Figure \ref{composite} shows the image composites from the three visits, with unadorned images in the top row and contoured images in the bottom row. The projected antisolar (yellow arrows) and negative heliocentric velocity vectors (white arrows) are superimposed.  The isophotes on Bennu show no evidence of preferential elongation in  either the antisolar or velocity directions, and there is no evidence of a dust trail in the data, even in 2023 September, where the in-plane observing geometry would maximize the surface brightness of any such a trail. The median apparent magnitudes of Bennu from the three visits vary from V = 22.86 (2012 September), 22.60 (2012 December) to 22.17 (2023 September), as a result of the changing observing geometry (Table \ref{geometry}).

\subsection{Time Resolved Nucleus Photometry}

We obtained photometry from the WFC3 images using a circular aperture of projected radius 0.2\arcsec~(5 pixels) with sky subtraction from the median signal within a contiguous annulus extending to an outer radius of 55 pixels (2.2\arcsec).  WFC3 data are afflicted with large numbers of cosmic rays.  We visually inspected each image for possible cosmic ray contamination and, where found, rejected the images from further consideration.  Four contaminated images were rejected.  Calibration was obtained by using the WFC3 Exposure Time Calculator\footnote{\url{https://etc.stsci.edu/etc/input/wfc3uvis/imaging/}} to note that a V = 0 sun-like (G2V) source gives a count rate 4.72$\times10^{10}$ s$^{-1}$.  This calibration is approximate, both because the F350LP filter bandpass is very broad compared to the V filter and because Bennu is slightly redder than the Sun in reflected light.  Nevertheless, the resulting average apparent magnitudes are within a few $\times$0.01 magnitudes of the JPL Horizons prediction and are of sufficient accuracy for our purposes.  The apparent magnitudes are listed in Table \ref{photometry}.

\cite{Her13} determined  the phase function of Bennu, finding a linear coefficient $\beta_V = 0.040\pm0.003$ magnitudes (degree)$^{-1}$ over the 15\degr~to 96\degr~phase angle range.  We applied this phase angle correction to the apparent magnitudes from Hubble and list the resulting absolute magnitudes in Table \ref{photometry}.  The average absolute magnitudes from the new data are compared with earlier values  in Table \ref{absmags}.  Uncertainties in the last column of the table include the errors of measurement but not uncertainty on the phase function. Given the large phase angles (up to $\sim$82\degr~in UT 2023 September) the latter are dominant; for example, at $\alpha$ = 82\degr, the 1$\sigma$ uncertainty in absolute magnitude due to the $\pm$0.003 magnitudes (degree)$^{-1}$ phase function uncertainty reaches 0.25 magnitudes.  The absolute magnitudes from Hubble are $\sim$0.25 magnitudes brighter than the value reported from ground-based data by \cite{Her13} but consistent with the updated value by \cite{Her19}.  It is not clear whether this difference simply reflects the phase function uncertainty or the existence of a color term in the F350LP to V filter conversion.

\begin{figure}
\epsscale{0.99}

\plotone{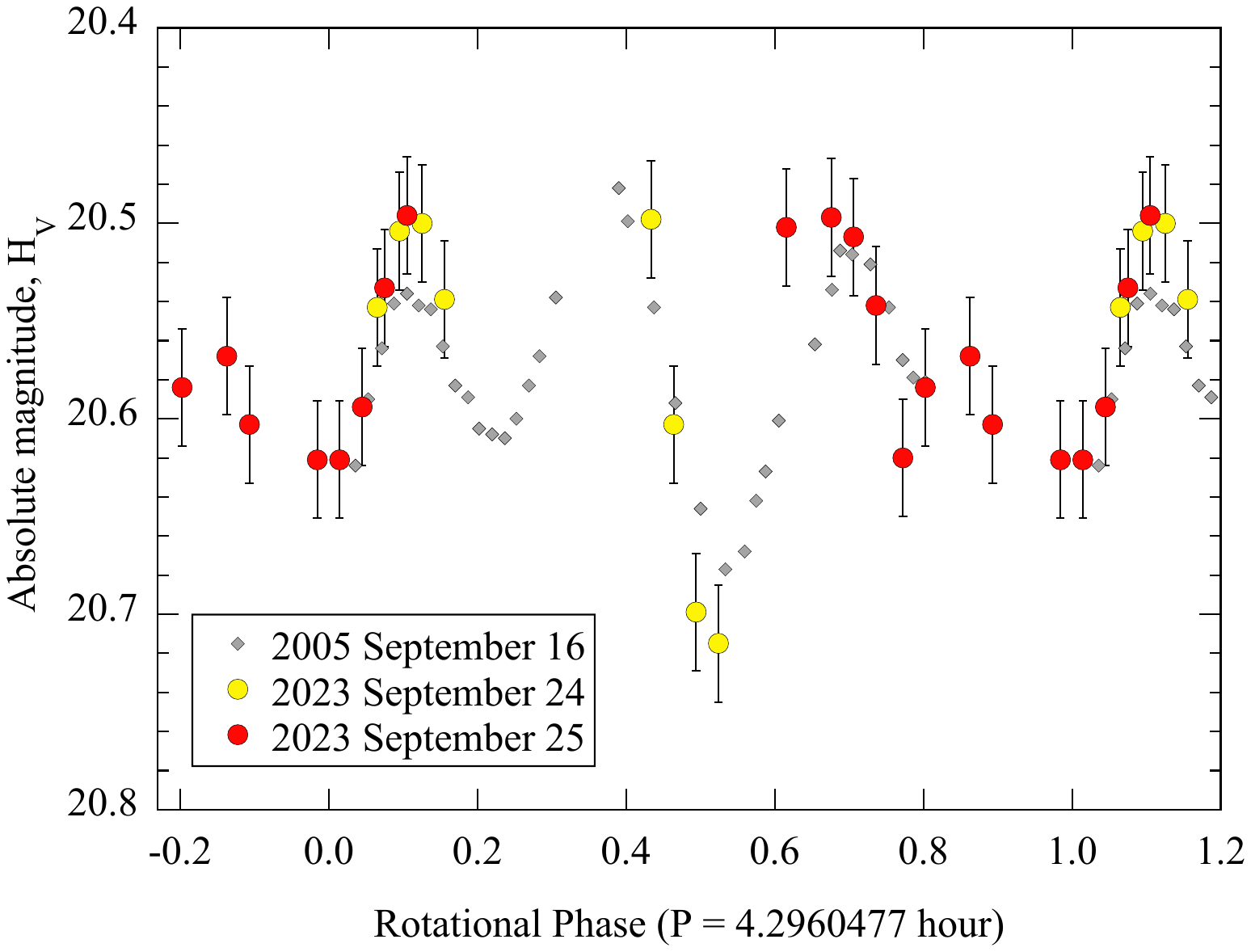}

\caption{Aligned lightcurves from UT 2023 September 24 (filled yellow circles) and 25 (red circles) compared with the lightcurve of UT 2005 September 14 (small gray diamonds) from \cite{Nol19}.   Error bars on the 2023 data are $1\sigma = \pm$0.03 magnitudes.   \label{lightcurves}}
\end{figure}

Figure \ref{lightcurves} shows the rotational lightcurve of Bennu phased from the HST data using a period $P$ = 4.2961 hour, compared with the lightcurve from 2005 September 16 measured by \cite{Her13} and taken from Figure 4 of \cite{Nol19}.  The latter has been shifted to best match the data from 2023.  Bennu's lightcurve has three maxima per rotation, rather than the two maxima expected from rotational symmetry, because it is influenced by small scale structures (including individual boulders) on the surface.  The photometric range, particularly near phase $\sim$0.5, appears smaller in the binned data from 2005 than in 2023, presumably as a result of changing viewing geometry.  Despite this, and with the exception of a single measurement from 2023 September 25 (at phase 0.63) the shapes of the two lightcurves are similar and there is no ambiguity in their alignment. Error bars on the 2023 data are estimated as 1$\sigma = \pm$0.03 magnitudes.  Error bars on the 2005 data are not plotted for clarity but, judging from the scatter in measurements repeated over four rotations, are likely to be at least as large. A preliminary analysis (Dr. Michael Nolan, private communication) indicates that the phase of the Bennu lightcurve supports the continuation of the spin acceleration reported by \cite{Nol19}.  We defer a detailed analysis of the spin state of Bennu to a later paper.

\subsection{Dust Coma}
\label{coma}

The escape velocity from Bennu varies from nearly zero at the equator (because of centripetal acceleration) to $\sim$0.2 m s$^{-1}$ at the poles \citep{Jia20}). Particles launched below the escape velocity can follow sub-orbital and bound trajectories  within a Hill radius, $r_{Hill}$, of the body.   The  Hill radius is 

\begin{equation}
r_{Hill} = q \left(\frac{\rho_B}{3\rho_{\odot}}\right)^{1/3}\frac{a_B}{R_{\odot}}
\end{equation}

\noindent where $\rho_B = 1190\pm15$ kg m$^{-3}$ and $a_B= 245$ m are the bulk density and effective radius of Bennu \citep{Lau19a}, $\rho_{\odot} = 1300$ kg m$^{-3}$ and $R_{\odot} = 7\times10^8$ m are the density and radius of the Sun, and $q = 0.897$ au is the perihelion distance.
Substitution gives $r_{Hill}$ = 32 km, about 130 times the radius of Bennu.  Numerical integrations suggest that from 1\% to 10\% of ejected particles larger than 1 $\mu$m  could temporarily populate the Hill sphere \citep{Jia20}. Unfortunately, even  at its closest approach to Earth ($\Delta$ = 0.551 au in 2023 September c.f.~Table\ref{geometry}), the Hill sphere radius subtends only 0.08\arcsec, comparable to the 0.08\arcsec~FWHM of WFC3 data. Therefore, our WFC3 data do not have the resolution to resolve material trapped within the Hill sphere of Bennu.

\begin{figure}
\epsscale{0.9}

\plotone{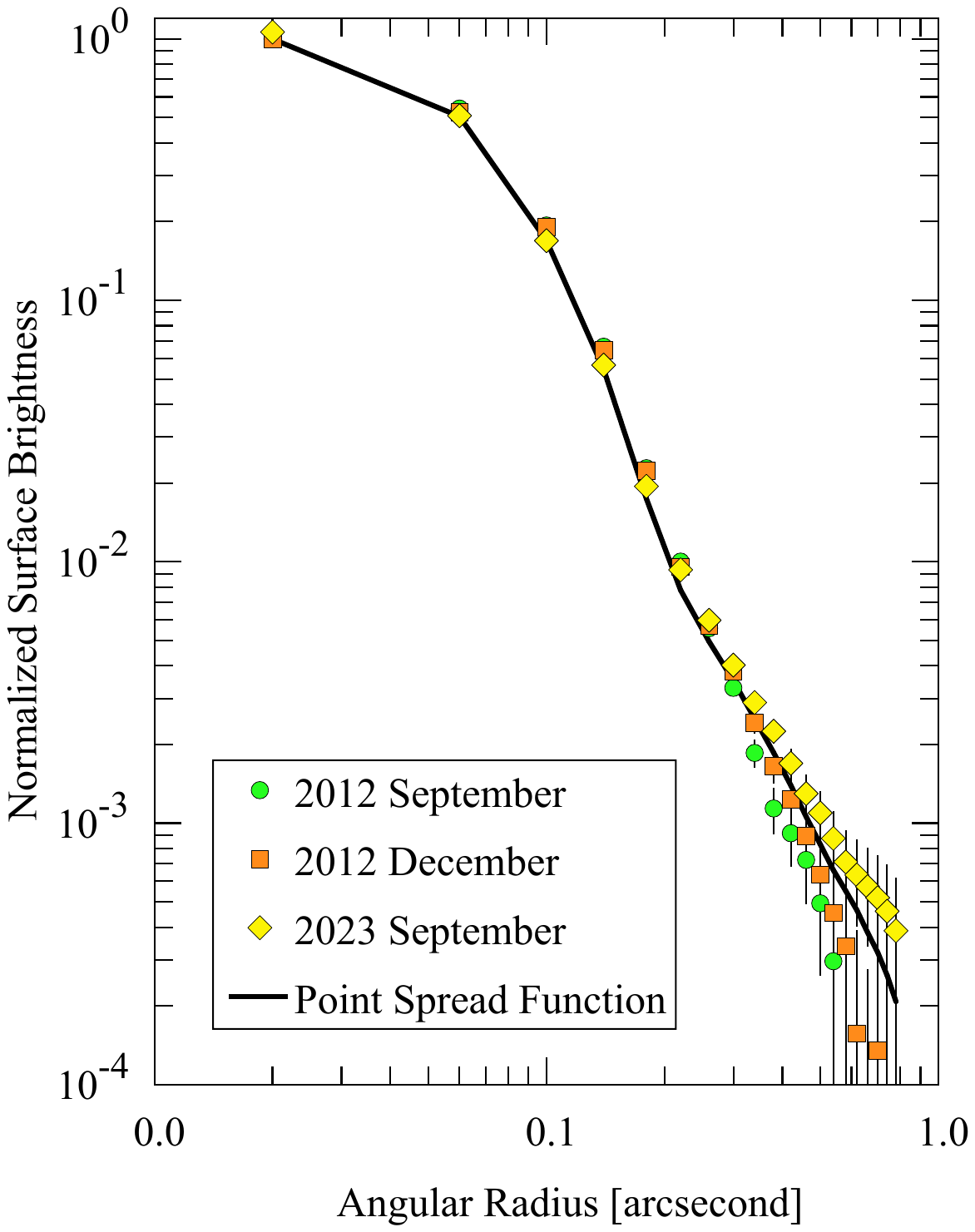}

\caption{Surface brightness profiles computed from concentric circular aperture photometry.  Error bars show estimated systematic sky background uncertainty 1$\sigma$ = 2$\times10^{-4}$ pixel$^{-1}$, estimated from spatially dispersed measurements of the residual sky level around Bennu. \label{profiles}}
\end{figure}

In this section we seek evidence for a near-nucleus dust coma beyond Bennu's Hill sphere.  Figure \ref{profiles} shows, $\Sigma(\theta)$, the surface brightness averaged within concentric, circular annuli, as a function of  angular distance from the photocenter, $\theta$, with sky subtraction from the median signal in a separate annulus having inner and outer radii 50 pixel and 150 pixel, respectively.  We show profiles for both months in 2012 and for 2023, with September 24 and 25 combined.  All three profiles are normalized at the central pixel for ease of comparison.  Peak surface brightnesses from the three epochs are $\Sigma(0\degr)$ = 18.32 magnitudes arcsec$^{-2}$ (2012 September), 18.05 magnitudes arcsec$^{-2}$ (2012 December), and 17.42 (2023 September) magnitudes arcsec$^{-2}$, with the differences being a result of the varying observational geometry (Table \ref{geometry}).  We estimated the sky background uncertainties, shown as 1$\sigma$ = 2$\times10^{-4}$  error bars in Figure \ref{profiles}, from the scatter in a series of local measurements within boxes 5$\times$5 pixels in size, distributed around the location of Bennu in the composite data. The uncertainties result from a combination of intrinsic errors in the flat field,  incomplete removal of overlapping trailed field stars and galaxies,  faint interference fringes present in some flat-fielded data and, especially, from a complex internal reflection field within the WFC3 optics.    The solid black line in Figure \ref{profiles} shows the HST point spread function (PSF), which we measured using reference images of sidereally guided images from the Mikulski Archive for Space Telescopes\footnote{\url{https://mast.stsci.edu/portal/Mashup/Clients/Mast/Portal.html}}.  The PSF shown in Figure \ref{profiles} was measured from stars in F350LP imaging data from GO 13767 (2015 May 31) and 15145 (2018 February 21) with exposure times of 485 s and 352 s, respectively.

The three profiles are in close agreement for $\Sigma(\theta)/\Sigma(0\degr) \ge 10^{-3}$, but at fainter levels the profiles begin to diverge from each other and from the PSF.  This divergence is consistent with the systematic uncertainty resulting from the background subtraction.  We measured a limit to the excess brightness in the 0.5\arcsec~to 1.0\arcsec~radius range as $<$1\% of the total light from Bennu.  Scaling from the mean projected cross-section of Bennu, $C = 0.19$ km$^2$, we set an upper limit to the dust cross-section as $C = 1900$ m$^2$ in this annulus.  


The measured ejection speeds of centimeter-sized particles, $U$, lie in the range 0.05 $\le U \le$ 3 m s$^{-1}$ \citep{Lau19b}.  In the following, we take the geometric mean, $U \sim$ 0.5 m s$^{-1}$, as a working estimate of the  speed of centimeter sized particles.  As is well known from the study of dust trails in the orbits of comets and asteroids, low velocity particles stay close to the orbit plane of the parent body and appear on the plane of the sky as thin, parallel-sided trails distinct from those of (high ejection velocity) comets.

The mass of an optically thin collection of spheres is related to their combined scattering cross-section, $C$, by 

\begin{equation}
M = 4\rho_B \bar{a} C/3
\label{mass}
\end{equation}

\noindent where $\rho_B$ is the particle density and $\bar{a}$ the mean particle size. In the absence of better evidence we assume that the ejected particles follow a differential power law distribution such that the number of particles with radii between $a$ and $a + da$ is $n(a) da = \Gamma a^{-q}da$, where $\Gamma$ and $q$ are constants.  With this assumption, and particle sizes in the range $a_{min} \le a \le a_{max}$, the cross-section weighted mean radius is given by

\begin{equation}
\bar{a} = \frac{\int_{a_{min}}^{a_{max}} a \pi a^2 n(a) da}{\int_{a_{min}}^{a_{max}}  \pi a^2 n(a) da}
\label{abar}
\end{equation}

The limits in the integrals of Equation \ref{abar} demand some consideration.  We set $a_{max} \sim$ 10 cm based on the largest particles detected in the OSIRIS-REx data \citep{Her19}. The choice of $a_{min}$ is more subtle.  Unfortunately, particles much smaller than a centimeter could not be detected by  OSIRIS-REx regardless of their abundance, owing to the limited sensitivity of the spacecraft camera.  While  sub-centimeter (even sub-micron) particles can contribute to the measured HST signal if they are present, small particles can also be selectively removed from a photometry annulus by solar radiation pressure. As a crude criterion by which to assess the importance of radiation pressure clearing, we compare the distance over which a particle can be accelerated with the width of the photometry annulus.
The latter, in meters, is $\Delta r = s \Delta (r_2 - r_1)$, where $r_2$ and $r_1$ are the outer and inner angular radii in arcseconds, respectively, and $s = 7.25\times10^5$ m (arcsec)$^{-1}$ au$^{-1}$ is the distance subtending 1\arcsec~from $\Delta$ = 1 au. 
In the absence of radiation pressure, the time taken for a particle to cross this distance is 

\begin{equation}
\tau \sim \frac{s \Delta (r_2 - r_1)}{U} \textrm{~~~~~[s]}
\label{tau}
\end{equation}

\noindent  where $U$ is the particle speed at ejection.  For example, with $\Delta$ = 0.55 au, $(r_2 - r_1)$ = 0.5\arcsec, and $U$ = 0.5 m s$^{-1}$, the crossing time is $\tau \sim 4\times10^5$ s (about 4.6 days).  During this time, the deflection by radiation pressure is

\begin{equation}
\ell = \frac{1}{2}\beta g_{\odot} \tau^2
\label{ell}
\end{equation}

\noindent where $\beta$ is the radiation pressure efficiency factor and $g_{\odot}$ is the local gravitational acceleration towards the Sun.  Combining  Equations \ref{tau} and \ref{ell} gives  

\begin{equation}
\ell = \frac{1}{2}\beta g_{\odot} \left(\frac{s \Delta (r_2 - r_1)}{U}\right)^2
\end{equation}

\noindent We consider that particles are deflected from the photometry annulus when $\ell \gg s \Delta (r_2 - r_1)$ and, from this inequality, we find

\begin{equation}
\beta \gg \frac{2 U^2}{s\Delta  (r_2 - r_1)g_{\odot}}.
\end{equation}

\noindent  Substituting $U$ = 0.5 m s$^{-1}$, $\Delta$ = 0.55 au, $(r_2 - r_1) =$ 0.5\arcsec, and $g_{\odot} = 7.0\times10^{-3}$ m s$^{-2}$ at $r_H$ = 0.92 au (c.f., Table \ref{geometry}), gives $\beta \gg 3.6\times10^{-4}$.  For dielectric spheres, $\beta$ and the particle radius, $a$, are related by $\beta \sim 10^{-6}/a$, with $a$ measured in meters \citep{Boh83}.  Therefore, we conclude that particles with $a \ll 2.8\times10^{-3}$ m (2.8 mm) would be depleted by being swept from the photometry annulus by radiation pressure faster than they could cross it, diminishing their contribution to the scattered light.   This is a crude estimate but suffices to show that submillimeter particles can be deflected on timescales shorter than the nominal aperture crossing time.   On this basis, we set $a_{min}$ = 2.8 mm.

\begin{figure}
\epsscale{0.99}

\plotone{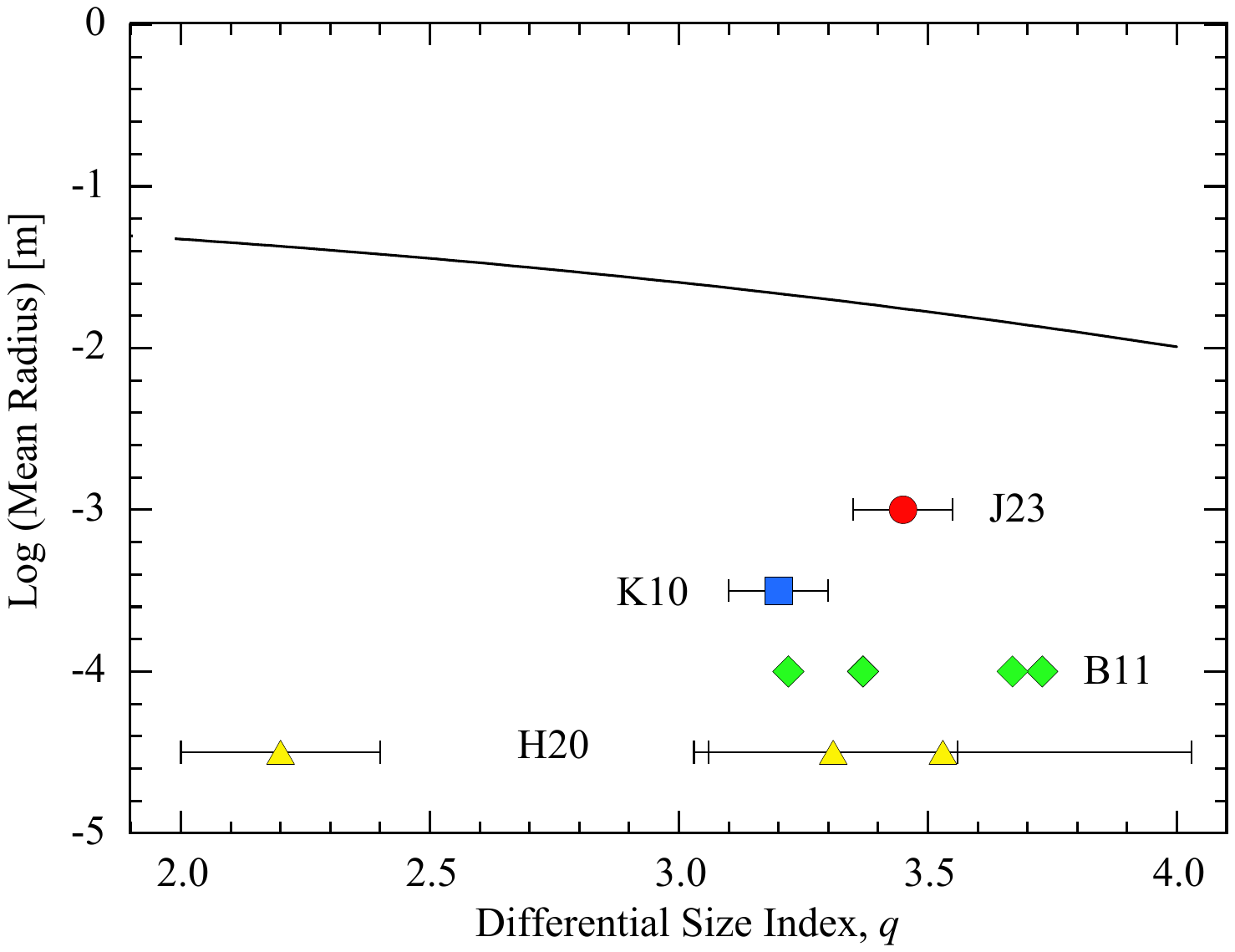}

\caption{Effective mean particle radius as a function of the size distribution index, from Equation \ref{abar}.  For comparison, we show  size index measurements from in-situ Bennu data (H20; \cite{Her20}),  meteor showers  (B11: \cite{Bla11}),  sungrazing comets  (K10: \cite{Kni10}) and  a disintegrated comet (J23:  \cite{Jew23}).    \label{abar_vs_q}}
\end{figure}

Resulting solutions to Equation \ref{abar} are plotted as a function of index $q$, assuming $a_{min}$ = 2.8 mm  and $a_{max}$ = 10 cm, in Figure \ref{abar_vs_q} (solid black line).  
For comparison, we also plot three size index determinations for Bennu centimeter-sized ejecta \citep{Her20} (yellow triangles), and measurements of millimeter-scale meteor shower particles \citep{Bla11} (green diamonds), decameter sized sungrazing comets \citep{Kni10} (blue square) and  submillimeter sized and larger particles released by the disintegration of long-period comet C/2021 A1 (Leonard) \citep{Jew23} (red circle).  (Note that the vertical position of the datapoints in the Figure is arbitrary). Other than a single  measurement, the size distributions all lie near $q \sim 3.5$, which we adopt as the appropriate index for Bennu here.    For $q$ = 3.5, Figure \ref{abar_vs_q} gives  $\bar{a} \sim$ 1.7 cm.\footnote{The size distribution of particles larger than 0.3 cm on the surface of Bennu  was found to have $q$ in the range from 3.3$\pm$0.1 to 4.0$\pm$0.2 \citep{Bur21}, bracketing the index we have adopted for ejected particles.  However, the ejection mechanism might well be size dependent, such that these indices are not necessarily applicable to the distribution of ejected material.} The figure also shows that the value of $\bar{a}$ is a weak function of $q$ across the plotted range and so the choice of size distribution index is not critical here.

Lastly, the dust mass production rate, $dM/dt \sim M/\tau$, from Equations \ref{mass} and \ref{tau}, can be written as

\begin{equation}
\frac{dM}{dt} \sim \frac{4\rho_B \bar{a} C_d U}{3s \Delta (r_2 - r_1)} \textrm{~~~~[kg } \textrm{s}^{-1}].
\label{dmbdt}
\end{equation}

\noindent As above, we take $\bar{a} = 1.7$ cm and $U$ = 0.5 m s$^{-1}$ based on particle data from OSIRIS-REx. We substitute $r_2$ = 1.0\arcsec, $r_1$ = 0.5\arcsec, finding $dM/dt \lesssim 0.12$ kg s$^{-1}$.  This is consistent with the upper limit to mass loss in September 2018, $dM/dt \lesssim$ 0.15 kg s$^{-1}$, deduced from in-situ measurements and using a different set of model assumptions \citep{Her19}.

\subsection{Dust Trail}
\label{trail}

Large dust particles  in comets and active asteroids  appear concentrated in thin, parallel-sided trails, morphologically distinct from the fan-shaped tails of (high ejection velocity) material (e.g., see examples in \cite{Jew12} and \cite{Jew24}).  Trails are especially evident  when observations are taken at a small angle from the projected orbital plane, $\delta_i$, as is the case for our 2023 data (Table \ref{geometry}). 

\begin{figure}
\epsscale{0.95}


\plotone{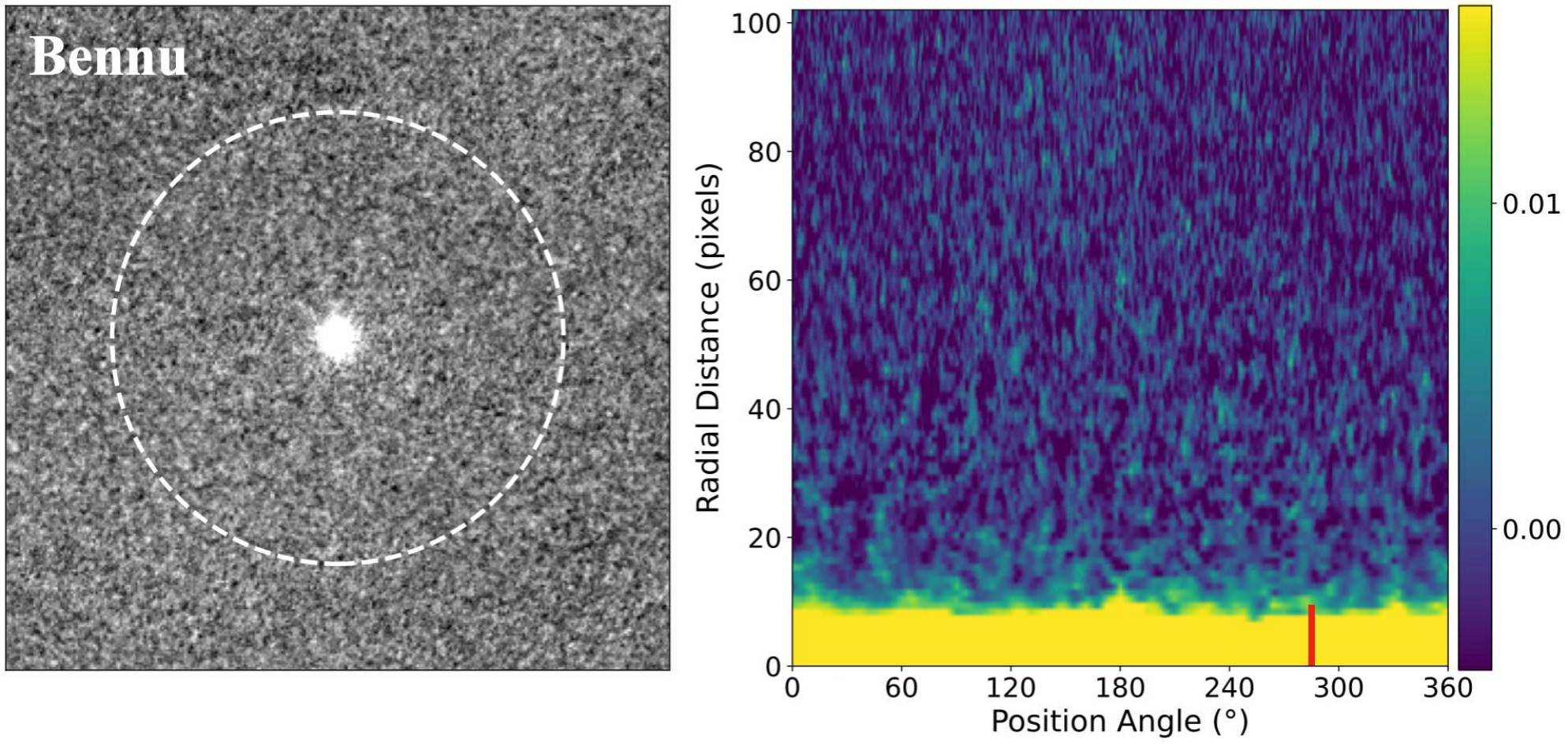}
\plotone{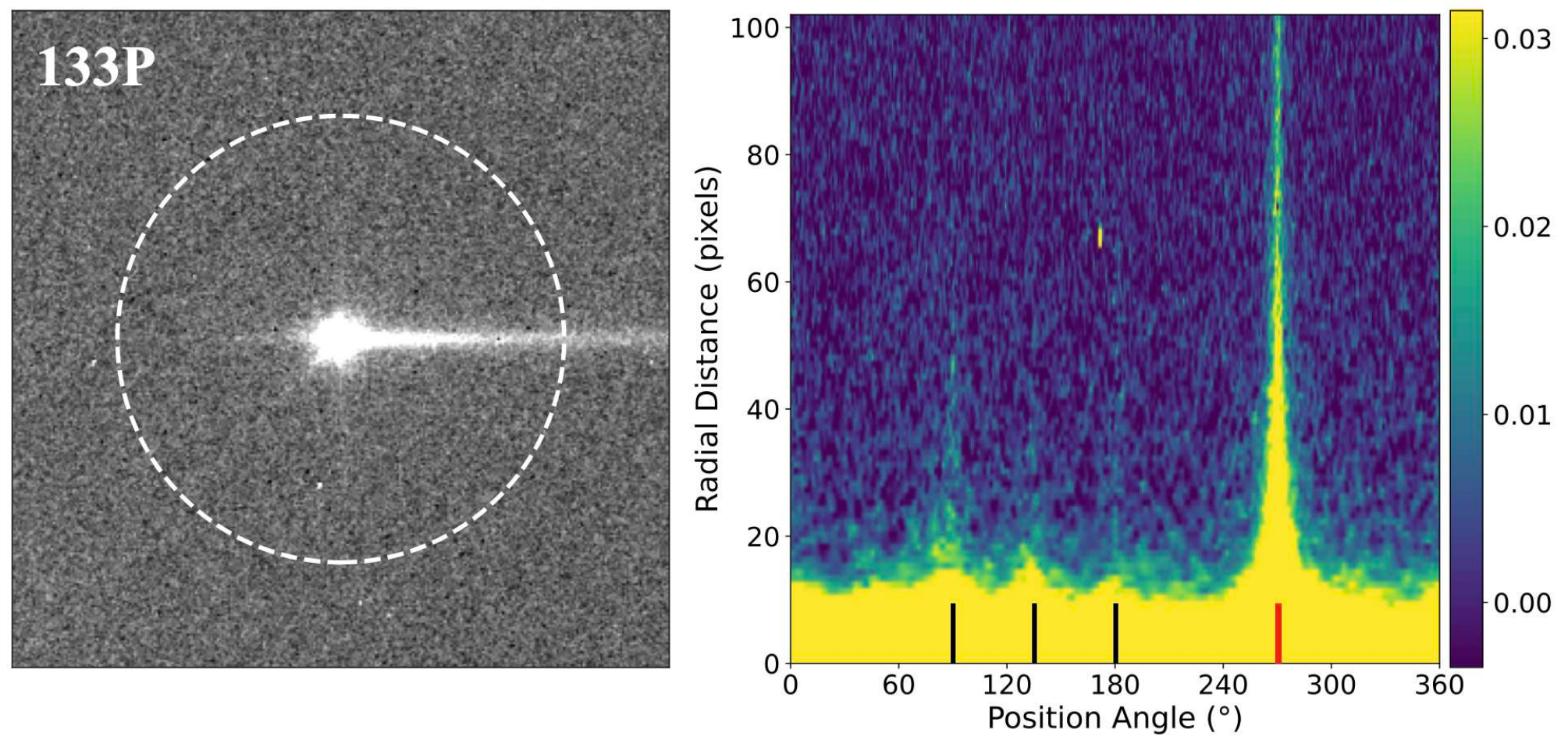}

\caption{(Upper:) Composite of 21 uncontaminated Hubble images of Bennu (total exposure 6090 s) at $r_H$ = 0.924 au on UT 2023 September 24/25.  No significant radial structures are evident.   (Lower:)  Same but for a composite image of 133P/Elst-Pizarro at $r_H$ = 2.725 au on UT 2013 July 10 having 3984 s total integration  (from \cite{Jew14}) with azimuthal plot on the right. The dotted circles in the left-hand panels for both objects have radius 100 pixels = 4.0\arcsec.  A vertical red line marks the projected negative heliocentric velocity vector.  Vertical black lines show significant radial image artifacts.    \label{radial}}
\end{figure}

We sought evidence for a dust trail associated with Bennu.  For this purpose, we transformed the image into polar coordinates centered on the Bennu photocenter, mapping the intensity at 1\degr~angular intervals over a range of radial distances up to 100 pixels (4\arcsec). A similar approach was used by \cite{Kwo25}. The resulting azimuthal intensity map (in ADU) is shown for the UT 2023 September 24/25 data as a function of position angle in the upper two panels of Figure \ref{radial}.  No excess emission is evident, and the expected trail position angle (negative heliocentric velocity is marked in the figure by a vertical red line) is conspicuously empty.  (Similarly, no evidence for a trail is present in the observations from 2012).    

For comparison, we show in the lower part of Figure \ref{radial}, the corresponding azimuthal profile of active asteroid 133P/Elst-Pizarro, also observed using WFC3 and HST \citep{Jew14}.  By comparison with Bennu, the trail in 133P is visually obvious, where it results from the presence of 0.2 to 10 mm sized particles following a $q \sim$ 3.25 power law with a total mass loss rate of order 1 kg s$^{-1}$.  

\begin{figure}
\epsscale{0.75}

\plotone{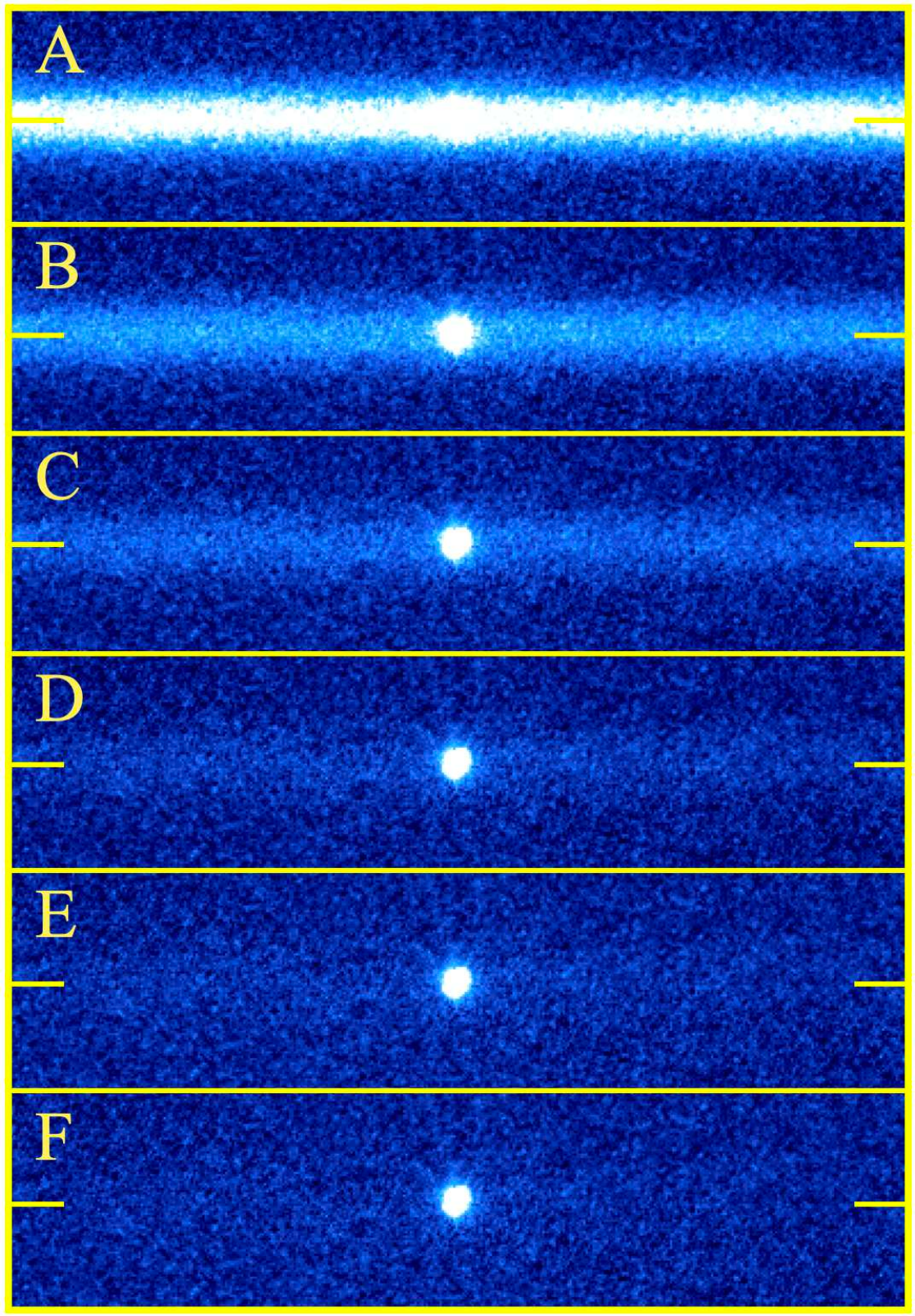}

\caption{Simulated trails on UT 2023 September 24/25 having, from top to bottom, peak surface brightnesses Panel A: 23.4, B: 24.1, C: 24.8, D: 25.5, E: 26.2 and F: 26.9 magnitudes (arcsecond)$^{-2}$.   Short horizontal lines to the left and right of each panel show the mid-lines of the digitally added trails. The panels are shown with a fixed stretch. The region shown in each panel is 16\arcsec~wide, centered on Bennu. \label{trails}}
\end{figure}

The visibility of a trail in optical data depends on the surface brightness and geometry of the trail, particularly on its width perpendicular to the projected orbit plane.  The trail of active asteroid 133P/Elst-Pizarro, for instance, has a perpendicular width $\sim$1\arcsec~when observed from the orbit plane of the body.  Accordingly, we simulated a Gaussian dust trail of 1\arcsec~width (FWHM) and having a range of peak surface brightnesses from $\Sigma$ = 23.4 magnitudes (arcsecond)$^{-1}$ to 26.9 magnitudes (arcsecond)$^{-1}$ (Figure \ref{trails}).   Based on the simulations, we conclude that a Bennu trail with $\Sigma \lesssim$ 26.5 magnitudes (arcsecond)$^{-1}$ would be evident in the  data, which we take as an estimate of the limiting sensitivity of the 2023 September data to diffuse emission.  

We integrated the signal perpendicular to the trail, then converted to the scattering cross-section of dust by scaling from the photometry of Bennu.   We find that the minimum detectable trail (c.f, Panel E in Figure \ref{trails}) has $C_T =$ 420 m$^2$ (arcsecond)$^{-1}$.   As in section \ref{coma}, the distance subtended by 1\arcsec~is $s\Delta$ and the time spent by a dust particle in crossing this distance is $\tau = s\Delta/U$, where $U$ is the particle speed.  The limit to the mass loss rate set by the non-detection of a trail is calculated using Equation \ref{dmbdt} with $U$ = 0.5 m s$^{-1}$, giving  $dM/dt < 5.8\times10^{-3}$ kg s$^{-1}$ for particles with $\bar{a}$ = 1 cm.  The limiting rate set by the absence of a trail is stronger than the rate set by measurement of the near-nucleus PSF because the trail has only a faint sky background whereas the wings of the PSF are affected by light from the nucleus.  

For comparison (Table \ref{massloss}), \cite{Her19} used spacecraft observations to set a limit to the production rate $dM/dt < 0.15$ kg s$^{-1}$, albeit in a limited 34 minute observational window. \cite{Eme14} used thermal infrared observations from the Spitzer Space Telescope to infer a maximimum dust mass  $M \lesssim 7\times10^5$ kg within 4750 km of the nucleus (see their Table 8). With $U$ = 0.5 ms$^{-1}$, this gives $dM/dt \lesssim$ 0.07 kg s$^{-1}$.  While all mass loss rate estimates are necessarily model dependent, it is clear that mass loss from Bennu is one to two orders of magnitude weaker than from 133P/Elst-Pizarro, for which $dM/dt \sim$ 1 kg s$^{-1}$ in its active state (\cite{Jew14}, \cite{Hsi26}).

\section{Discussion}
Given the empirical mass loss limits  summarized in Table \ref{massloss},  the Hubble trail limit ($dM/dt < 0.006$ kg s$^{-1}$) is the most sensitive of those based on direct observations. However, the Bennu rates all fall far short of the mass loss rates measured in active asteroids (where peak values are $\sim$1 kg s$^{-1}$) and in comets (where rates $\sim10^3$ kg s$^{-1}$ are common at 1 au), consistent with the absence of measurable coma in Bennu.   Indeed, Bennu is only known to be active because of the exquisite sensitivity provided by in-situ, as opposed to remote, observations.  This raises the obvious possibility that many or most asteroids might be  similarly active at low levels, but have escaped detection owing to the limited sensitivity of ground-based data. Assuming this to be the case, we next estimate the total rate of production of debris across the asteroid belt to be expected at the Bennu rate.

\cite{Her20} considered the bursty nature of the mass loss events in Bennu and concluded that, while the short-term value can be much higher, the orbit-averaged mass loss rate in centimeter sized and larger particles is $dM/dt \sim10^{-7}$ kg s$^{-1}$. If so, the specific mass loss rate from Bennu, represented as a sphere of radius $r_n$ = 245 m, is $\mathcal{L} =  (dM/dt)/(4\pi r_n^2)$.  Substitution gives $\mathcal{L} \sim 10^{-7}$ kg km$^{-2}$ s$^{-1}$.

We crudely estimate the cumulative mass production to be expected if all asteroids larger than radius $r_{min}$ lose mass at the same specific rate,  $\mathcal{L}$.  This cumulative rate is $dM/dt(r_{min}) = C_T(r_{min})\mathcal{L}$, where $C_T(r_{min})$ is the sum of the surface areas of all asteroids larger than radius $r_{min}$.  
To calculate $C_T$ we represent the main-belt asteroids, of total mass $M = 2.4\times10^{21}$ kg \citep{Pit18}, as a collection of spheres with radii distributed as a $q$ = 3.5 power law between minimum and maximum radii $r_{min}$ and $r_{max}$, respectively, with $r_{max} \gg r_{min}$.  The total (as opposed to projected) surface area is then given by

\begin{equation}
C_T = \frac{3M}{\rho (r_{min} r_{max})^{1/2}}
\end{equation}

\noindent where we also take $\rho$ = 1190 kg m$^{-3}$ (the same as Bennu) as the average bulk density of the asteroids, and $r_{max}$ = 500 km, the radius of the largest asteroid, (1) Ceres.  

By substitution, the resulting asteroid-wide production rate is 

\begin{equation}
\frac{dM}{dt} \sim \frac{850}{r_{min}^{1/2}}  \textrm{~~~~~[kg}~\textrm{s}^{-1}].
\label{DMBDT}
\end{equation}

\noindent Equation \ref{DMBDT} clearly gives only an order of magnitude estimate and it would be unwise to over-interpret its significance.  The choices of $q$ and of $r_{min}$ are unclear, for example, because the size distribution of the smallest asteroids is not well measured and their number is likely to be depleted  (through collisions, and through rotational disruption under YORP torques) relative to a $q$ = 3.5 power law.  At the large size end of the distribution, the escape of impact-produced particles will be limited by the gravity of the body.  Both effects suggest that $dM/dt$ from this simple treatment is over-estimated.  Nevertheless, it is safe to conclude that for any $r_{min} \gtrsim$ 1 m, the active asteroids produce dust and debris at a rate at most 1\% to 10\% of the $\sim10^4$ to 10$^5$ kg s$^{-1}$ rate needed to maintain the Zodiacal cloud in steady state \citep{Nes11}.

\begin{figure}
\epsscale{0.99}

\plotone{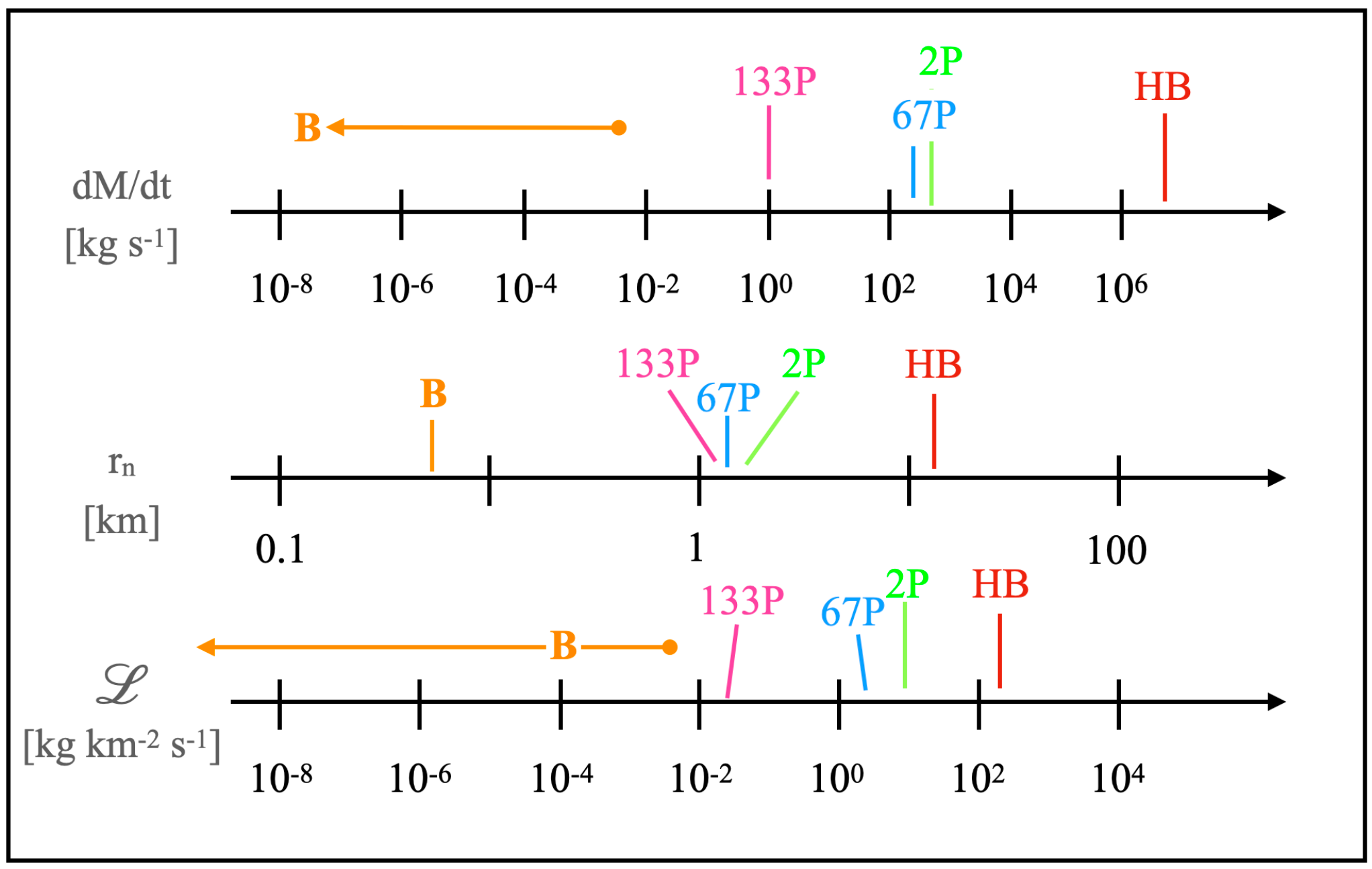}

\caption{Comparison of (top)  mass loss rates, $dM/dt$ [kg s$^{-1}$], (middle) nucleus radii, $r_n$ [km], and (bottom)  specific mass loss rates, $\mathcal{L}$ [kg km$^{-2}$ s$^{-1}$], for Bennu (B) and cometary bodies 133P/Elst-Pizarro, 2P/Encke,  67P/Churyumov-Gerasimenko and C/1995 O1 (Hale-Bopp). The Bennu rates are those derived from HST observations described in the present work.
\label{comparisons}}
\end{figure}

Lastly, we compare the specific mass loss rate from Bennu (from HST data) with the same quantity in other  objects in Figure \ref{comparisons}.  There, in addition to the above-mentioned range of possible values for Bennu (marked B in the figure), we show values for the active asteroid 133P/Elst-Pizarro (radius and mass loss rate 1.9 km, 1 kg s$^{-1}$, respectively), short-period comets 2P/Encke (2.4 km, 2000 kg s$^{-1}$) and 67P/Churyumov-Gersimenko (2.5 km, 600 kg s$^{-1}$, and long-period comet C/1995 O1 (Hale-Bopp) (32 km, 2$\times10^6$ kg s$^{-1}$).  For simplicity, no attempt has been made to correct the different mass loss measurements to a common heliocentric distance.   Neither do we account for observations showing that peak activity levels can vary from orbit to orbit in any given object.  The figure shows that the dimensions of the plotted bodies span a range of about two orders of magnitude, while the specific mass loss rates span a much more substantial range of order 10$^4$ to 10$^5$.

\section{SUMMARY}

We used the Hubble Space Telescope to search for mass loss from active asteroid (and former spacecraft target) (101955) Bennu from a vantage point close to its orbital plane. We use the observations to set upper limits to mass in ejected particles in the mm to cm size range.  We find that

\begin{itemize}

\item The first limit, obtained from the near-nucleus surface brightness profile, corresponds to an upper limit to the mass loss rate $dM/dt <$ 0.12 kg s$^{-1}$.  

\item The second limit, based on the absence of a low surface brightness large particle trail, is $dM/dt < 6\times10^{-3}$ kg s$^{-1}$.  

\item These observational limits to the activity are one to two orders of magnitude smaller than mass loss rates measured in other active asteroids. 

\item Bennu's orbitally averaged specific mass loss rate, $\mathcal{L}$ = 10$^{-7}$ kg km$^{-2}$ s$^{-1}$, if applicable to the surfaces of all asteroids, predicts a cumulative mass production rate $dM/dt \sim 850 r_{min}^{-1/2}$ kg s$^{-1}$.  With $r_{min} >$ 1 m, this is $\lesssim$1\% to 10\% of the rate needed to sustain the Zodiacal dust cloud in steady state.

\end{itemize}

\acknowledgments
We thank Jessica Agarwal, Jane Luu and the anonymous referee for comments on  this paper. Based on observations made with the NASA/ESA Hubble Space Telescope, obtained from the data archive at the Space Telescope Science Institute. STScI is operated by the Association of Universities for Research in Astronomy, Inc. under NASA contract NAS 5-26555. The HST data presented in this article were obtained from the Mikulski Archive for Space Telescopes (MAST) at the Space Telescope Science Institute. The datasets on which this work is based can be found at \dataset[doi: 10.17909/z2xd-xn62]{https://doi.org/10.17909/z2xd-xn62}. Support for this work was provided by NASA through grants GO 17215 and AR 17030 from the Space Telescope Science Institute, which is operated by auRA, Inc., under NASA contract NAS 5-26555.



\clearpage


\begin{deluxetable}{lllllllllll}
\tabletypesize{\scriptsize}
\tablecaption{Journal of Observations \label{geometry}}
\tablewidth{0pt}
\tablehead{
\colhead{GO} & \colhead{UT Date} & Exp\tablenotemark{a} & $r_H$ \tablenotemark{b} & $\Delta$ \tablenotemark{c}&  \colhead{$\alpha$\tablenotemark{d}}   &  \colhead{$\theta_{-\odot}$\tablenotemark{e}}  & \colhead{$\theta_{-V}$\tablenotemark{f}}  &  \colhead{$\delta_i$\tablenotemark{g}} & \colhead{$\nu$ \tablenotemark{h}}  }
\startdata
 13118 & 2012 Sep 17/18 	 & 40$\times$305 & 1.216 & 0.961 & 53.4 & 99.9 & 282.1 & 0.77 & 236.5 \\
 13118  & 2012 Dec 10/11   & 40$\times$305 & 0.952 & 0.962 & 61.9 & 71.3 & 232.5 & -6.02 & 311.2 \\
 17215 & 2023 Sep 24 & 10$\times$290 & 0.924 & 0.550 & 81.6 & 283.6 & 283.1 & 0.14 & 325.8 \\
 17215 & 2023 Sep 25 & 15$\times$290 & 0.923 & 0.551 & 81.6 & 283.8 & 283.6 & 0.06 & 326.3 \\

\enddata

\tablenotetext{a}{ Number of exposures $\times$ exposure time in seconds}
\tablenotetext{b}{ Heliocentric distance in au.}
\tablenotetext{c}{ Geocentric distance in au.}
\tablenotetext{d}{ Phase angle in degree}
\tablenotetext{e}{ Position angle of the projected anti-solar direction}
\tablenotetext{f}{ Position angle of the projected negative heliocentric velocity vector}
\tablenotetext{g}{ Angle of the Earth above the projected orbital plane}
\tablenotetext{h}{ True anomaly, degree}
\end{deluxetable}


\clearpage


\begin{deluxetable}{llccc}
\tablecaption{Mean Magnitudes \label{photometry}}
\tablewidth{0pt}
\tablehead{
\colhead{N} &\colhead{Universal Date\tablenotemark{a}}  & V [mag]\tablenotemark{b} & H [mag]\tablenotemark{c}  }
\startdata 
1	&	September 24.893	&	22.16	$\pm$ 0.03	&	20.54	\\				
2	&	September 24.898	&	22.12	$\pm$ 0.03	&	20.50	\\				
3	&	September 24.904	&	22.12	$\pm$ 0.03	&	20.50	\\				
4	&	September 24.909	&	22.16	$\pm$ 0.03	&	20.54	\\				
5	&	September 24.959	&	22.12	$\pm$ 0.03	&	20.50	\\				
6	&	September 24.964	&	22.22	$\pm$ 0.03	&	20.60	\\				
7	&	September 24.970	&	22.32	$\pm$ 0.03	&	20.70	\\				
8	&	September 24.975	&	22.34	$\pm$ 0.03	&	20.72	\\				
9	&	September 25.020	&	22.24	$\pm$ 0.03	&	20.62	\\				
10	&	September 25.025	&	22.20	$\pm$ 0.03	&	20.58	\\				
11	&	September 25.036	&	22.19	$\pm$ 0.03	&	20.57	\\				
12	&	September 25.041	&	22.22	$\pm$ 0.03	&	20.60	\\				
13	&	September 25.349	&	22.12	$\pm$ 0.03	&	20.50	\\				
14	&	September 25.360	&	22.12	$\pm$ 0.03	&	20.50	\\				
15	&	September 25.366	&	22.13	$\pm$ 0.03	&	20.51	\\				
16	&	September 25.371	&	22.16	$\pm$ 0.03	&	20.54	\\				
17	&	September 25.416	&	22.24	$\pm$ 0.03	&	20.62	\\				
18	&	September 25.421	&	22.24	$\pm$ 0.03	&	20.62	\\				
19	&	September 25.426	&	22.21	$\pm$ 0.03	&	20.59	\\				
20	&	September 25.432	&	22.15	$\pm$ 0.03	&	20.53	\\				
21	&	September 25.437	&	22.12	$\pm$ 0.03	&	20.50	\\				
\enddata 

\tablenotetext{a}{ Universal date of the observation in 2023}
\tablenotetext{b}{ Apparent V magnitude and 1$\sigma$ uncertainty}
\tablenotetext{c}{Absolute magnitude computed from V assuming phase function 0.04 magnitudes degree$^{-1}$. The statistical uncertainty on H is the same as on V but the absolute uncertainty could be several $\times$0.1 magnitude as a result of the phase correction.}
\end{deluxetable}

\clearpage


\begin{deluxetable}{ccccc}
\tablecaption{Mean Magnitudes \label{absmags}}
\tablewidth{0pt}
\tablehead{
\colhead{UT Date} &\colhead{ID}  & FILTER &  mag\tablenotemark{a} & H \tablenotemark{b}   }
\startdata
2012 Sep 17 & GO 13118 & F350LP & 22.80$\pm$0.01	& 20.32$\pm$0.16  \\
2012 Dec 10 & GO 13118 & F350LP& 22.54$\pm$0.01 	& 20.25$\pm$0.21  \\

2023 Sep 24 & GO 17215 & F350LP& 22.20$\pm$0.03 	& 20.40$\pm$0.24  \\
2023 Sep 25 & GO 17215 & F350LP& 22.18$\pm$0.01 	& 20.38$\pm$0.24  \\
\hline
\cite{Her13} & N/A & V & -- & 20.61$\pm$0.20 \\
\cite{Her19} & N/A & V & -- & 20.41$\pm$0.02 \\

\hline

\enddata 

\tablenotetext{a}{ Mean apparent magnitude within a 0.8\arcsec~radius aperture}
\tablenotetext{b}{ Mean absolute magnitude computed assuming phase function 0.040$\pm$0.003 magnitudes (degree)$^{-1}$ \citep{Her13}.  The quoted uncertainty is due to the combined measurement and phase angle corrections, but does not include scatter from incomplete sampling of the lightcurve at each epoch.}

\end{deluxetable}


\clearpage
\clearpage


\begin{deluxetable}{lclclllllll}
\tablecaption{Mass Loss Rates \label{massloss}}
\tablewidth{0pt}
\tablehead{
 \colhead{UT Date\tablenotemark{a}} & $r_H$\tablenotemark{b} & $dM/dt$ \tablenotemark{c} & Method\tablenotemark{d} & Source \tablenotemark{e}  }
\startdata
2007 May 8     & 1.138 &  $<$0.07  & Spitzer thermal imaging & \cite{Eme14} \\
2018 Sep 11/12 & 1.208 &  $<$0.15   & In-situ optical imaging& \cite{Her19} \\
2023 Sep 24/25 & 0.923 &  $<$ 0.12 & Hubble, coma & This Work,  \S \ref{coma}  \\
2023 Sep 24/25 & 0.923 &  $<$ 0.006 & Hubble, trail &  This Work, \S \ref{trail} \\
\hline
Orbit Average & -- & $\lesssim10^{-7}$ & -- & \cite{Her20} \\

\enddata

\tablenotetext{a}{ UT Date of the measurement}
\tablenotetext{b}{ Heliocentric distance in au.}
\tablenotetext{c}{ Mass loss rate in kg s$^{-1}$}
\tablenotetext{d}{ Method of measurement}
\tablenotetext{e}{ Reference}

\end{deluxetable}



\clearpage 


%
%

\clearpage 


%


\clearpage 


\clearpage


\end{document}